\documentstyle[]{mn}

\title[Orbital periods of 22 sdB stars]{Orbital periods of
  twenty two subdwarf B stars}

\author[L.\, Morales-Rueda et al.] {L.\,Morales-Rueda$^{1}$,
  P.\,F.\,L.\,Maxted$^{2,1}$, T.\,R.\,Marsh$^{1}$,
  R.\,C.\,North$^{1,3}$, U.\,Heber$^{4}$ \\
  $^{1}$Department of Physics and Astronomy, University of
  Southampton, UK (lmr@astro.soton.ac.uk, trm@astro.soton.ac.uk)\\
  $^{2}$School of Chemistry and Physics, Keele University, UK
  (pflm@astro.keele.ac.uk)\\
  $^{3}$Meteorological Office, London Road, Bracknell, Berkshire,
  RG12 2SZ, UK\\
  $^{4}$Dr Remeis-Sternwarte, Astronomisches Institut der Universit{\"
  a}t Erlangen-N{\" u}rnberg, Sternwarstrasse 7, 96049 Bamberg,
  Germany\\
  }

\date{Accepted 0000 000 00; Received 0000 000 00; in original
  form 0000 000 00}

\pagerange{\pageref{firstpage}--\pageref{lastpage}}

\def\LaTeX{L\kern-.36em\raise.3ex\hbox{a}\kern-.15em
    T\kern-.1667em\lower.7ex\hbox{E}\kern-.125emX}

\begin{document}

\newcommand{\gk}{\mbox{GK~Per}} 
\newcommand{\etal}{\mbox{et\ al.}}
\newcommand{\kmsec}{\,\mbox{$\mbox{km}\,\mbox{s}^{-1}$}}
\newcommand{\phispin}{$\phi_{\tiny spin}$}
\newcommand{\phiorb}{$\phi_{\tiny orb}$}
\newcommand{\msun}{\hbox{$\hbox{M}_\odot$}}
\newcommand{\ha}{\hbox{$\hbox{H}\alpha$}}
\newcommand{\hb}{\hbox{$\hbox{H}\beta$}}
\newcommand{\hgam}{\hbox{$\hbox{H}\gamma$}}
\newcommand{\heii}{\hbox{$\hbox{He\,{\sc ii}\,$\lambda$4686\,\AA}$}}
\newcommand{\hei}{\hbox{$\hbox{He\,{\sc i}\,$\lambda$4472\,\AA}$}}
\newcommand{\ciiiniii}{\hbox{$\hbox{C\,{\sc iii}/N\,{\sc
        iii}\,$\lambda\lambda$4640--50\,\AA}$}}

\label{firstpage}

\maketitle

\begin{abstract}
  Subdwarf B (sdB) stars are thought to be core helium burning stars
  with low mass hydrogen envelopes. In recent years it has become
  clear that many sdB stars lose their hydrogen through interaction
  with a binary companion and continue to reside in binary systems
  today. In this paper we present the results of a programme to
  measure orbital parameters of binary sdB stars. We determine the
  orbits of 22 binary sdB stars from 424 radial velocity measurements,
  raising the sample of sdBs with known orbital parameters to 38. We
  calculate lower limits for the masses of the companions of the sdB
  stars which, when combined with the orbital periods of the systems,
  allow us to discuss approximate evolutionary constraints. We find
  that a formation path for sdB stars consisting of mass transfer at
  the tip of the red giant branch followed by a common envelope phase
  explains most, but not all of the observed systems. It is
  particularly difficult to explain both long period systems and short
  period, massive systems. We present new measurements of the
  effective temperature, surface density and surface helium abundance
  for some of the sdB stars by fitting their blue spectra. We find
  that two of them (PG0839+399 and KPD1946+4340) do not lie in the
  Extreme Horizontal Branch (EHB) band indicating that they are
  post-EHB stars.
\end{abstract}


\section{Introduction}

Subdwarf B (sdB) stars can be identified with models for Extreme
Horizontal Branch (EHB) stars. The surface gravities and temperatures
of sdB stars suggest that they have helium cores of mass $\sim
0.5\,\msun$ and thin hydrogen envelopes of mass $\leq 0.02\,\msun$
\cite{h84,s94}. A recent asteroseismological study of an sdB star
results in a value for its mass of 0.49$\pm$0.02\,$\msun$ \cite{b01}.
Several evolutionary scenarios have been proposed to explain the
formation of sdB stars, in particular the loss of the hydrogen
envelope. Evolution within a binary star is an effective method for
envelope removal, and yet it is hard to see why this should have
happened to a horizontal branch star since it would have been much
larger during its preceding red giant stage. A solution to this
problem was presented by D'Cruz et al.\ \shortcite{dc96} who found
that if a red giant star with a degenerate helium core loses its
hydrogen envelope when it is within $\sim$0.4\,magnitudes of the tip
of the red giant branch, the core can go on to ignite helium, despite
the dramatic mass loss, and may then appear as an sdB star. The
advantage of this model is that it very nicely explains the masses of
sdB stars as a consequence of the core mass at the helium flash.
D'Cruz et al.\ \shortcite{dc96} supposed that mass loss occurred
because of an enhancement of the stellar wind, but it could as well
have been driven by binary interaction.

If sdB stars do form within binary systems and if they still have
their companions, then the companions must be low-mass main-sequence
stars or compact stellar remnants to avoid outshining the sdB star. If
so, it is probable that in many cases the companions were unable to
cope with the mass transferred from the sdB progenitor and a single
``common'' envelope formed around the two stars. Driving off such
envelopes drains energy and angular momentum from the binary orbit,
which as a result becomes much smaller than it was at the start of
mass transfer \cite{w84}. It is therefore possible that many sdB stars
are now members of close binary systems. Maxted et al.\
\shortcite{m01} found exactly this, discovering 21 binary sdB stars in
a sample of 36, suggesting, after allowance for detection efficiency,
that some two-thirds of all sdB stars are in short period binary
systems ($P \la 10\,\mathrm{d}$). The other third seems to be made up
a combination of long period binary stars that avoided a common
envelope phase \cite{gls00} and apparently single sdB stars.

If D'Cruz et al.'s \shortcite{dc96} model of the formation of sdB
stars is correct, then the stage immediately prior to mass transfer in
binary sdBs is well defined. This, together with the fact that the
binary does not have enough time to change its orbital period
significantly following its emergence from the common envelope, makes
the sdB stars a superb population for testing models of the common
envelope phase. Moreover, the detection of sdB binary stars is not
compromised by the strong and poorly understood selection effects that
plague other populations of close binary stars, such as the
cataclysmic variable stars. The properties of sdB binaries (e.g.
their orbital period distribution) can be compared fairly directly
with the results of binary population synthesis codes and are
therefore a strong test of population synthesis models for binary
stars.

Following on from the detection of many binary stars by Maxted et al.
\shortcite{m01}, we started a project to measure their orbits. The
orbit of one of the new binary stars has been presented in Maxted et
al.\ \shortcite{m02a}. In this paper we present the orbits of a
further 22 systems.  We then consider the implications of the known
sdB binary stars for their evolution. It should be noted that our
sample is biased against sdB stars with G/K-type companions as the
majority of our stars were selected from the PG survey which excludes
most stars that show a Ca{\sc ii} H-line.

\section{Observations and Reduction}

The data used in this study were taken with the Intermediate
Dispersion Spectrograph (IDS) at the 2.5m Isaac Newton Telescope (INT)
on the island of La Palma. Two different configurations of the IDS
were used for the observations. The first setup consisted of the 500
mm camera with the R1200R grating centred in \ha\ and the TEK (1kx1k)
Charge couple device (CCD) giving a dispersion of 0.37\,\AA/pix and a
resolution of 0.9\,\AA. The second setup used the 235 mm camera with the
R1200B grating centred at 4350\AA\ and the thinned EEV10 (2kx4k) CCD
covering the Bamer lines from \hb\ to H$\epsilon$, giving a dispersion
of 0.48\,\AA/pix and a resolution of 1.4\,\AA. We carried out the
observations during six different runs. The dates when the
observations were taken and the setup used during each campaign are
given in Table~\ref{obs:joo}. We took two consecutive observations of
each object and bracketed them with CuAr plus CuNe frames to calibrate
the spectra in wavelength. We subtracted from each image a constant
bias level determined from the mean value in its over-scan region.
Tungsten flatfield frames were obtained each night to correct for the
pixel to pixel response variations of the chip. Sky flatfields were
also obtained to correct for the pixel to pixel variations of the chip
along the slit.  After debiasing and flatfielding the frames, spectral
extraction proceeded according to the optimal algorithm of Marsh
\shortcite{m89}. The arcs were extracted using the profile associated
with their corresponding target to avoid systematic errors caused by
the spectra being tilted.  Uncertainties on every point were
propagated through every stage of the data reduction.

\begin{table}
\caption{Journal of observations. Setup 1 is: INT + IDS + 500 mm + R1200R +
  $\lambda_c$=\ha. Setup 2 is: INT + IDS + 235 mm + R1200B +
  $\lambda_c$=4350\AA.}
\label{obs:joo}
\begin{center}
\begin{tabular}{llc}
Dates & Setup & \# of RV observations\\
\hline
10 -- 21 Apr 2000& 1 & 85\\  
5 -- 6 Feb 2001& 1 & 28\\
8 -- 13 Mar 2001& 1 & 130\\
1 -- 8 May 2001& 1 & 99\\
6 -- 11 Aug 2001& 2 & 56\\
27 Sep -- 6 Oct 2001& 2 & 26\\
\end{tabular}
\end{center}
\end{table}

\section{Results}

\subsection{Radial velocity measurements}
\label{results:rv}

\begin{figure*}
\begin{picture}(100,0)(10,20)
\put(0,0){\includegraphics{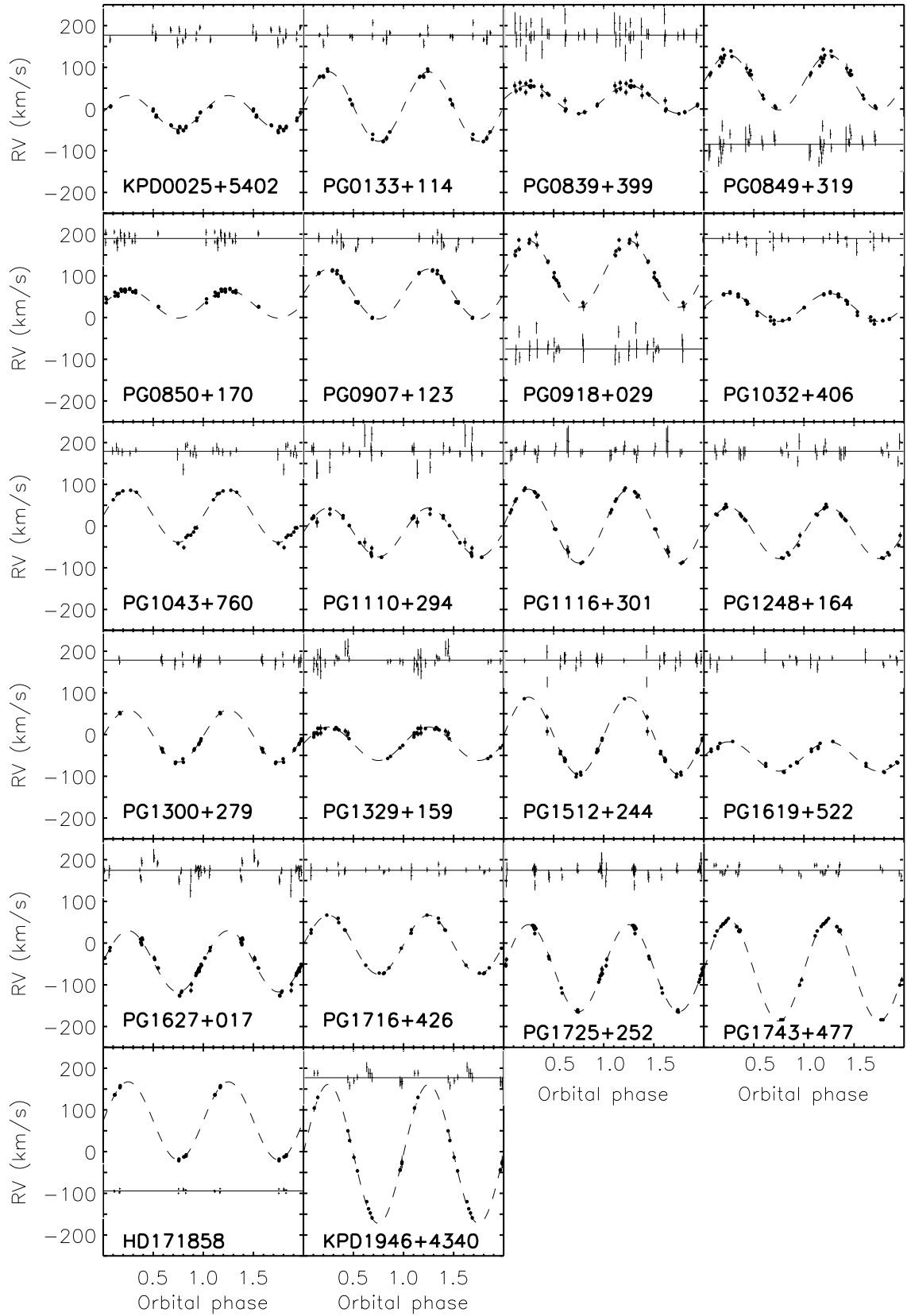}}
\noindent
\end{picture}
\vspace{225mm}
\caption{Each panel presents the radial velocity curve measured for
  each object. The data have been folded on the orbital period in each
  case. See table \ref{results:rv:tab2} for the list of periods,
  radial velocity semiamplitudes and systemic velocities. Included in
  each panel is a plot of the residuals to the fit. The vertical scale
  on which the residuals have been plotted is 10 times larger than the
  scale on which the radial velocities are plotted.}
\label{results:rv:rvtotal}
\end{figure*}

\begin{table*}
\caption{List of the orbital periods measured for the 22 sdBs
studied. T$_{0}$, the systemic velocity, $\gamma$, the radial velocity
semi-amplitude, K, the reduced $\chi^2$ achieved for the best alias,
the 2nd best alias and the $\chi^2$ difference between the 1st and 2nd
aliases are also presented. The number of data points used to
calculate the orbital period is given in the final column under n. See
text for a comment on the orbital period of PG1043+760.}
\label{results:rv:tab2}
\begin{center}
\begin{tabular}{lllccllll}
Object & HJD (T$_{0}$) & Period (d)& $\gamma$ (km/s) & K (km/s) &
 $\chi^{2}_{reduced}$ &2nd best alias (d)& $\Delta \chi^{2}$ & n \\
 & $-$2450000 & & & & & & \\
\hline

KPD0025+5402 & 2159.386(9) & 3.571(1) & $-$7.8$\pm$0.7
& 40.2$\pm$1.1 & 2.82 & 3.832(2) & 21 & 22\\

PG0133+114 & 2158.682(2) & 1.2382(2) & 6.0$\pm$1.0 & 83.2$\pm$0.8 &2.13
& 4.277(1) & 129 & 18\\

PG0839+399 & 1914.06(6) & 5.622(2) & 23.2$\pm$1.1 &
33.6$\pm$1.5 & 1.22& 4.720(1) & 16 & 24\\

PG0849+319 & 1841.992(3) & 0.74507(1) & 64.0$\pm$1.5 &
66.3$\pm$2.1 & 1.94& 0.426983(6) & 22 & 20\\

PG0850+170 & 1834.3(2) & 27.81(5) & 32.2$\pm$2.8 &
33.5$\pm$3.1 & 1.44& 13.86(2) & 13 & 23\\

PG0907+123 & 1840.62(3) & 6.1163(6) & 56.3$\pm$1.1 &
59.8$\pm$0.9 & 1.13&5.0619(5) & 30 & 16\\

PG0918+029 & 1842.310(4) & 0.87679(2) &
104.4$\pm$1.7 & 80.0$\pm$2.6 & 1.87&0.82644(3) & 53 & 18\\


PG1032+406 & 1888.66(2) & 6.779(1) & 24.5$\pm$0.5 &
33.7$\pm$0.5 & 1.63& 6.034(1) & 16 & 24\\

PG1043+760 & 1842.4877(7) & 0.1201506(3) &
24.8$\pm$1.4 & 63.6$\pm$1.4 & 1.62& 0.572097(5) & 40 & 14\\

PG1110+294 & 1840.49(3) & 9.415(2) & $-$15.2$\pm$0.9 &
58.7$\pm$1.2 & 1.50& 1.16397(3) & 32 & 21\\

PG1116+301 & 1920.834(2) & 0.85621(3) &
$-$0.2$\pm$1.1 & 88.5$\pm$2.1 & 0.52& 4.5237(8) & 26 & 16\\

PG1248+164 & 1959.853(4) & 0.73232(2) &
$-$16.2$\pm$1.3 & 61.8$\pm$1.1 & 0.93&0.688431(7) & 21 & 16\\

PG1300+279 & 1908.310(7) & 2.2593(1) & $-$3.1$\pm$0.9
& 62.8$\pm$1.6 & 0.65& 1.50254(4) & 26 & 16\\

PG1329+159 & 1840.579(1) & 0.249699(2) &
$-$22.0$\pm$1.2 & 40.2$\pm$1.1 & 0.92& 0.199694(2) & 17 & 23\\

PG1512+244 & 1868.521(2) & 1.26978(2) &
$-$2.9$\pm$1.0 & 92.7$\pm$1.5 & 0.81& 0.363261(1) & 37 & 20\\

PG1619+522 & 1837.0(1) & 15.357(8) & $-$52.5$\pm$1.1 &
35.2$\pm$1.1 & 1.38& 0.1153123(3) & 23 & 14\\

PG1627+017 & 2001.267(1) & 0.829226(8) &
$-$43.7$\pm$0.5 & 73.6$\pm$0.9 & 2.63& 0.836541(8) & 413 & 32\\

PG1716+426 & 1915.806(5) & 1.77732(5) &
$-$3.9$\pm$0.8 & 70.8$\pm$1.0 & 0.78& 2.62356(6) & 75 & 13\\

PG1725+252 & 1901.3977(8) & 0.601507(3) &
$-$60.0$\pm$0.6 & 104.5$\pm$0.7 & 1.08& 0.594906(3) & 512 & 30\\

PG1743+477 & 1921.1183(7) & 0.515561(2) &
$-$65.8$\pm$0.8 & 121.4$\pm$1.0 & 1.44&1.024201(6) & 58 & 18\\

HD171858 & 2132.241(6) & 1.529(8) & 73.8$\pm$0.8 &
93.6$\pm$0.7 & 0.65& 0.6352(8) & 28 & 12\\

KPD1946+4340 & 2159.0675(5) & 0.403739(8) &
$-$5.5$\pm$1.0 & 167.0$\pm$2.4 & 1.49&0.400780(8) & 78 & 14\\

\end{tabular}
\end{center}
\end{table*}

To measure the radial velocities we used least squares fitting of a
model line profile. The model line profile is the summation of three
Gaussian profiles with different widths and depths. For any given
star, the widths and depths of the Gaussians are optimised and then
held fixed while their velocity offsets from the rest wavelengths of
the lines in question are fitted separately for each spectrum; see
Maxted, Marsh \& Moran \shortcite{m00c} for further details of this
procedure.  For the data taken on the April 2000, the February, March,
and May 2001 observing runs, the fitting was performed to the \ha\ 
line whereas for the August and September 2001 observing campaigns,
the fitting was carried out simultaneously to all the Balmer lines
present in the spectra.  Table \ref{results:rv:tab} gives a list of
the radial velocity measurements for the 22 objects presented in this
paper.

Using the measured radial velocities of the lines, we determined the
orbital periods of our targets. The description of the orbital period
determination is given in Section~\ref{results:periods}. The results
of folding the radial velocities of each object on its orbital period
are plotted in Fig.~\ref{results:rv:rvtotal}. The error bars on the
radial velocity points are, in most cases, smaller than the size of
the symbol used to display them. For that reason we also display the
residuals of the fit on a scale 10 times larger than the scale of the
radial velocity curves. We find that there is no sign of ellipticity
in any of the radial velocity curves, not even at long periods where
departures from circular orbits might be expected. This is good
evidence for the action of the common envelope. The values of the
orbital periods, systemic velocities and radial velocity
semi-amplitudes for each system are given in
Table~\ref{results:rv:tab2}.

\subsection{Determination of orbital periods}
\label{results:periods}

We use a ``floating mean'' periodogram to determine the periods of our
targets (e.g. Cumming, Marcy \& Butler 1999). The method consists in
fitting the data with a model composed of a sinusoid plus a constant
of the form:
\[  \gamma + {\rm K}  \sin (2 \pi f (t - t_0)),\]
where $f$ is the frequency and $t$ is the observation time. The key
point is that the systemic velocity is fitted at the same time as $K$
and $t_0$.  This corrects a failing of the well-known Lomb-Scargle
\cite{l76,s82} periodogram which starts by subtracting the mean of the
data and then fits a plain sinusoid; this is incorrect for small
numbers of points.

We obtain the $\chi^2$ of the fit as a function of $f$ and then
identify minima in this function. Section~\ref{results:prob} gives a
detailed explanation on the probability of the periods obtained being
incorrect.

\begin{figure*}
\begin{picture}(100,0)(10,20)
\put(0,0){\includegraphics{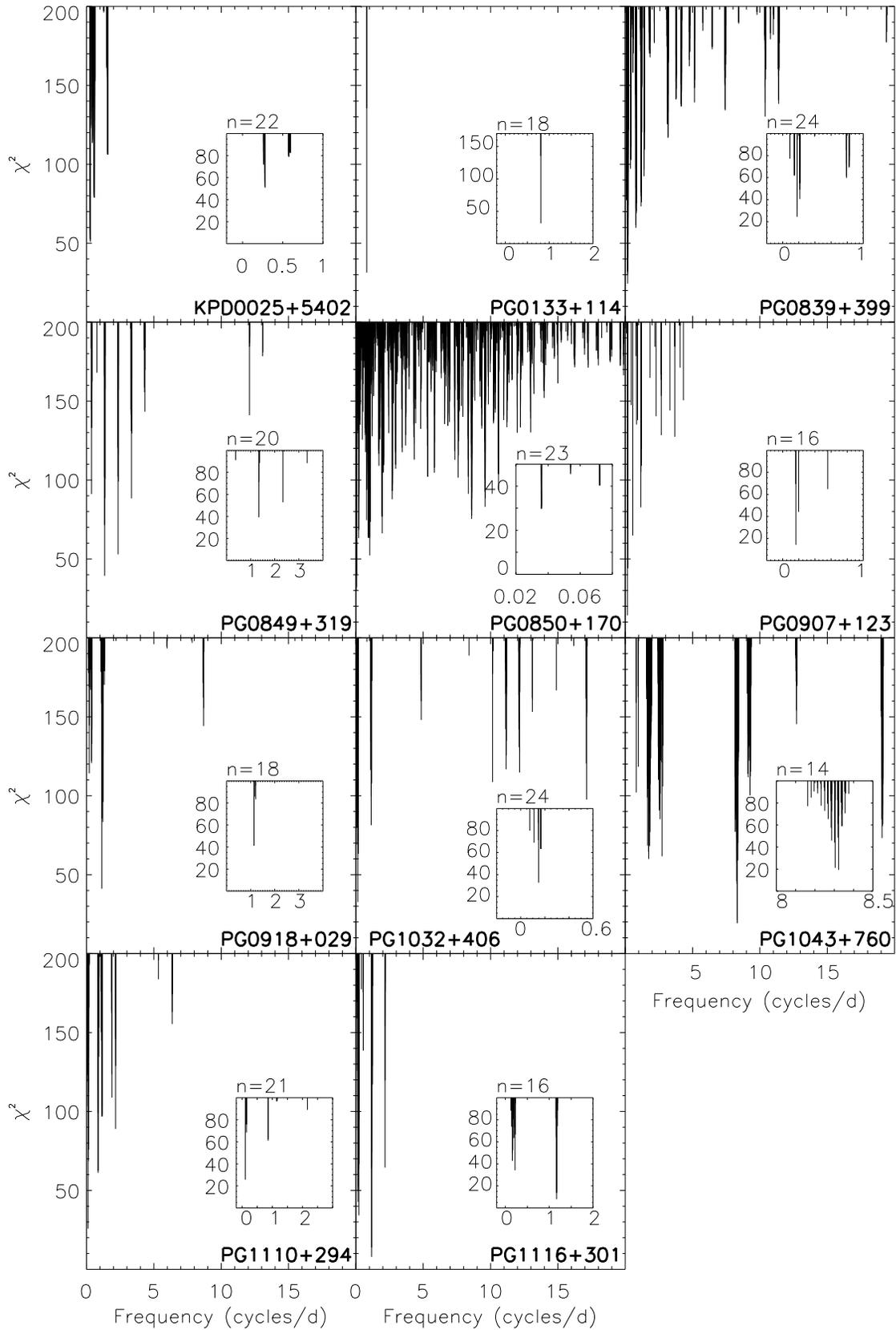}}
\noindent
\end{picture}
\vspace{225mm}
\caption{Each panel presents $\chi^2$ versus cycles/day obtained after
the period search was carried out. The frequency with the smallest
$\chi^2$ corresponds to the orbital frequency of the system. For
clarity we have also included an inset showing a blow up of the region
where the best period is. The number of radial velocity measurements
used for the period search calculations, n, is shown in each panel.}
\label{results:periods:pgram1}
\end{figure*}

\begin{figure*}
\begin{picture}(100,0)(10,20)
\put(0,0){\includegraphics{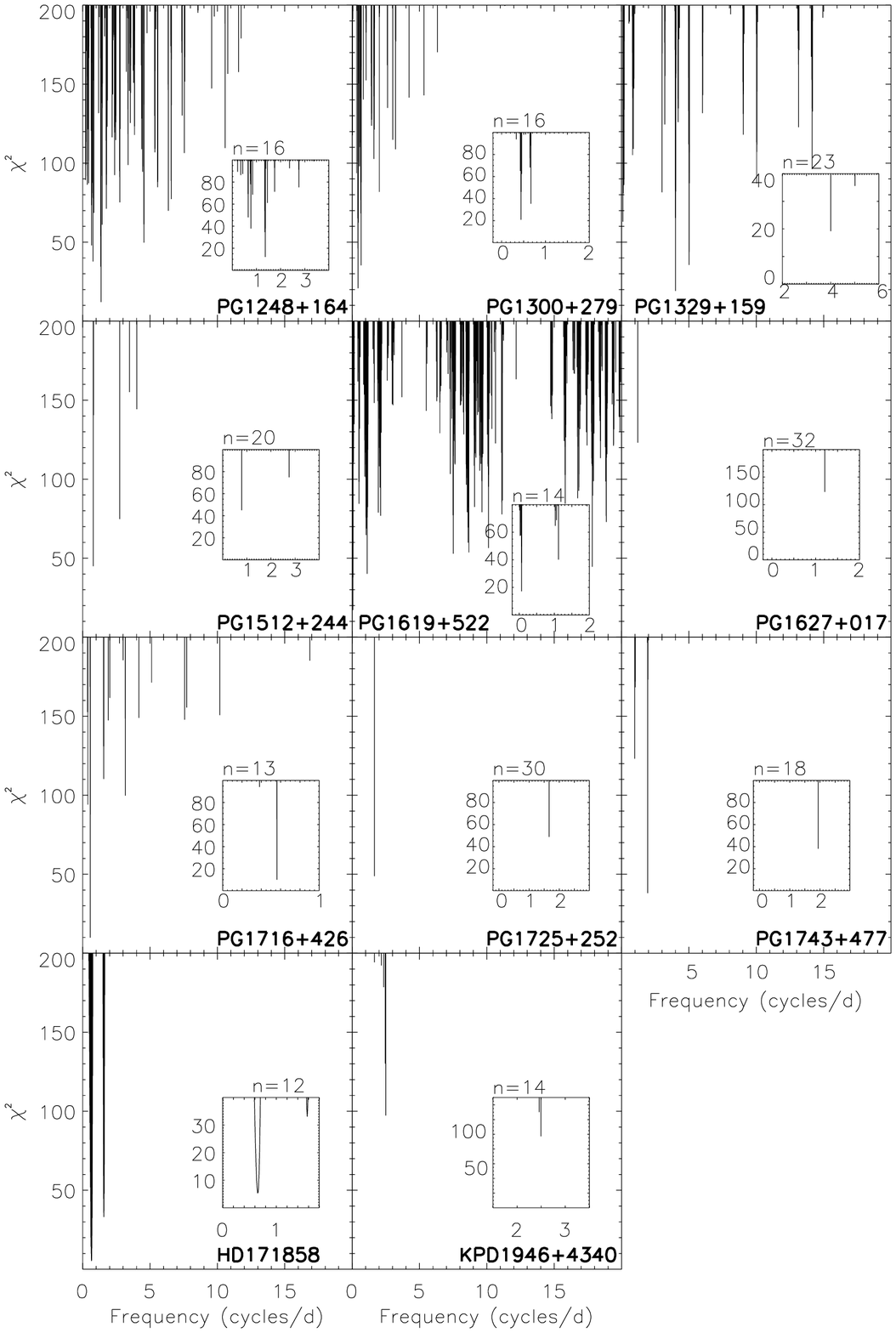}}
\noindent
\end{picture}
\vspace{225mm}
\caption{Same as in Fig.~\ref{results:periods:pgram1} for the remaining
  11 systems.}
\label{results:periods:pgram2}
\end{figure*}

Table~\ref{results:rv:tab2} gives a list of the orbital parameters
derived for each sdB binary star. The orbital period of the second
best period is also given, along with the difference in $\chi^2$
between the two best periods found. The resulting periodograms
($\chi^2$ versus orbital frequency) are given in
Figs.~\ref{results:periods:pgram1} and \ref{results:periods:pgram2}.
Each panel includes an blow up of the region in frequency where the
minimum $\chi^2$ is found.  It is clear from the figures that in all
cases, apart from PG1043+760, the difference in $\chi^2$ between the
best and the second alias is at least 10. We have made an exception
for PG1043+760, because in this case the competing aliases are so
close (owing to 1 cycle/year aliasing) that it makes more sense to
consider the nearest competing group of aliases.

\subsection{The probability that our periods are incorrect}
\label{results:prob}

A sometimes frustrating characteristic of radial velocity work is that
while one can very soon know for sure that a star is binary -- perhaps
after just two measurements -- it can take much longer to pin down the
orbital period. One somehow has to know when the orbital period is
``correct''. To compound this problem, another feature of radial
velocity orbits is that if one picks an incorrect alias, the period
can be \emph{completely} wrong, even when the quoted uncertainty on
the best-fit period is tiny. This is simply because the statistics are
not Gaussian so that an error of 100 or even 1000 times quoted
uncertainty can happen. Perhaps the most common way around this issue
is the ``method of overkill'' where one takes so much data as to put
the issue beyond any doubt, but this is necessarily inefficient.  This
is a particular problem with the sdB stars where there is no shortage
of potential targets, but always a shortage of telescope time in which
to observe them. Our approach whilst observing was to use the
rule-of-thumb that an orbit was determined once the best-fit orbit
improved upon the next-best by at least 10 in $\chi^2$. The basis for
this is that the probability of a period in the Bayesian sense is
dominated by the term $\exp -\chi^2/2$ (see the appendix), and so a
difference of more than 10 shows that the second-best alias is at
least $\exp 5 \approx 150$ times less probable than the best. This
rule-of-thumb was propagated into the submitted version of the paper,
but the referee made two cogent criticisms of this approach. First,
while the peak of the second alias may be $> 150$ times less probable
than the peak of the best alias, there is no guarantee that the total
probability of \emph{any} other period was as low. Second, some of our
$\chi^2$ values were larger than expected given the number of data
points, suggesting some extra source of uncertainty that if included
would reduce one's faith in the best alias.
 
Both of these criticisms were justified, and the first in particular
prompted us to develop a more rigorous approach which we describe
below. To address the second point we have computed the level of
systematic uncertainty that when added in quadrature to our raw error
estimates gives a reduced $\chi^2 = 1$. The reasoning behind this is
that there may be an un-accounted source of error such as true
variability of the star or slit-filling errors causing the poor fits
of a few stars. It seems unlikely that such errors will be either
correlated with the orbit or with the statistical errors we estimate,
and therefore we add a fixed quantity in quadrature with our
statistical errors as opposed to applying a simple multiplicative
scaling to them. In all cases we use a minimum value of $2\,{\rm
  km}\,{\rm s}^{-1}$ corresponding to 1/10$^{\rm th}$ of a pixel which
we believe to be a fair estimate of the true limits of our data. 

The end result is that in two cases there is indeed a relatively large
probability that our periods are in error (KPD0025+5402; PG1032+406),
nevertheless we continue to list them in this paper on the bases that
(a) the best periods remain strongly favoured in all cases and
therefore are probably correct and (b) the IDS spectrograph is being
withdrawn from service and it is not clear when we will get the
opportunity to gather more data. These systems (KPD0025+5402,
PG1032+406) therefore come with a health warning: there is a larger
than desirable probability that we may not have picked the best alias
for them. 

In most cases the probability of the orbital period being further than
1 and 10 per cent from our favoured value is the same. This is because
all the significant probability lies within a very small range around
the best period, with all the significant competition (i.e. next best
alias) placed outside the 10 per cent region around the best alias.
In the case of HD171858, the probability of the true orbital period
being further than 1 per cent from the one given in this paper is very
high. This is due to the short baseline of our observations (1 day)
which means that the period is not determined anything like as well as
the others and in particular is not confined to within 1 per cent of
the best fit. For the purpose of comparing observations to
evolutionary models there is no need for better than 10 per cent
measurements and we are quite certain that the period is correct
within those limits. The probabilities for all systems are listed in
Table~\ref{tab:probs} where we give the logarithm (base 10) of the
chance that the true period lies further than 1 and 10 per cent of our
favoured value.
 
\begin{table}
\caption{List of probabilities that the true orbital period of a
  system lies further than 1 and 10 per cent from our favoured
  value given in Table~\ref{results:rv:rvtotal}. Numbers quoted are
  the logs in base 10 of the probabilities. Column number 4 gives the
  value of the systematic uncertainty that has been added in
  quadrature to the raw error to give a $\chi^2$ that lies above the
  2.5 per cent probability in the $\chi^2$ distribution.}
\label{tab:probs}
\begin{center}
\begin{tabular}{lllc}
Object & 1\%\ & 10\%\ & systematic error\\
 & & & (km\,s$^{-1}$) \\
\hline
KPD0025+5402  & $-$1.32 & $-$2.29 & 5 \\
PG0133+114    & $-$15.80 & $-$15.80 & 3\\
PG0839+399    & $-$3.73 & $-$3.73 & 2\\
PG0849+319    & $-$4.19 & $-$4.19 & 3\\
PG0850+170    & $-$3.05 & $-$3.39 & 2\\
PG0907+123    & $-$5.04 & $-$5.04 & 2\\
PG0918+029    & $-$4.66 & $-$9.16 & 3\\
PG1032+406    & $-$2.03 & $-$2.03 & 3\\
PG1043+760    & $-$4.74 & $-$4.74 & 2\\
PG1110+294    & $-$6.66 & $-$6.66 & 2\\
PG1116+301    & $-$4.75 & $-$4.75 & 2\\
PG1248+164    & $-$4.60 & $-$4.92 & 2\\
PG1300+279    & $-$5.70 & $-$5.70 & 2\\
PG1329+159    & $-$3.57 & $-$3.57 & 2\\
PG1512+244    & $-$7.26 & $-$7.26 & 2\\
PG1619+522    & $-$5.24 & $-$5.24 & 2\\
PG1627+017    & $-$55.94 & $-$68.89 & 4\\
PG1716+426    & $-$5.17 & $-$5.17 & 2\\
PG1725+252    & $-$104.88 & $-$124.42 & 2\\
PG1743+477    & $-$26.45 & $-$26.45 & 2\\
HD171858      & $-$0.28 & $-$6.88 & 2\\
KPD1946+4340  & $-$12.96 & $-$15.32 & 2\\
\end{tabular}
\end{center}
\end{table}

To calculate these probabilities, we integrated equation~A4 from the
appendix of Marsh, Dhillon \&\ Duck \shortcite{mdd95}. Full details of
our method are given in the appendix to this paper. The calculation is
Bayesian, and therefore involves prior probabilities over all
parameters, i.e.  the systemic velocity, semi-amplitude, phase and
period.  These ``priors'' are uncertain and therefore the final
probabilities listed in Table~\ref{tab:probs} are similarly uncertain.
This we believe is the main reason why one would like the probability
of being incorrect to be \emph{very} small, so as to overcome any
plausible uncertainty in the prior probabilities; this is one
justification for the method of overkill. For most of our sample, this
is in fact the case, but we feel that the uncertainties are large
enough to leave a grain of doubt when the probability listed in
Table~\ref{tab:probs} rises above $0.1$\%, or $-3$ in the log, as it
does for the two stars discussed.

\section{Discussion}

\subsection{Effective temperature, surface gravity and helium abundance}

For those sdBs that we observed in the blue, we measured the effective
temperature, T$_{\rm eff}$, the surface gravity, $\log g$, and the
helium abundance, $\log (\rm He/H)$. We used Saffer et al.'s
\shortcite{s94} procedure to fit the profiles of the Balmer, the
He\,{\sc i} and the He\,{\sc ii} lines present in the spectra by a
grid of synthetic spectra. The synthetic spectra obtained from
hydrogen and helium line blanketed NLTE atmospheres \cite{n97} were
matched to the data simultaneously. For stars cooler than 27\,000\,K
we used the metal line-blanketed LTE model atmospheres of Heber, Reid,
\&\ Werner \shortcite{hrw00}. Before the fitting was carried out, we
convolved the synthetic spectra with a Gaussian function to account
for the instrumental profile. See Heber et al. \shortcite{hrw00} for
details of the models. The values of T$_{\rm eff}$, $\log g$, and
$\log (\rm He/H)$ for some of the objects are the same as those
presented by Maxted et al.\ \shortcite{m01}. Those that are different
from the ones obtained by Maxted et al. \shortcite{m01} and those for
which T$_{\rm eff}$ and $\log g$ have not been calculated previously
(KPD0025+5402 and KPD1946+4340) are given in
Table~\ref{dis:tefflogg:tab}.

All the objects from Table~\ref{dis:tefflogg:tab} apart from two lie
in or near the band defined by the zero-age extreme horizontal branch
(ZAEHB), the terminal-age extreme horizontal branch (TAEHB) and the He
main sequence (HeMS) and are therefore extreme horizontal branch stars
(EHB). (See Fig. 2 of Maxted et al.\ \shortcite{m01} for a $\log g$
versus T$_{\rm eff}$ plot for the objects not included in
Table~\ref{dis:tefflogg:tab}.) The two objects that lie outside the
EHB band are post-EHB stars. PG0839+399 appears in Saffer et al.\ 
\shortcite{s94} as an EHB star, and KPD1946+4340 is a new post-EHB
star.

\begin{table}
\caption{T$_{\rm eff}$, $\log g$ and $\log (\rm He/H)$ calculated for
  newly discovered sdB binaries (KPD0025+5402 and KPD1946+4340). We
  also present these values for other systems when they do not
  coincide with previously published measurements \protect\cite{m01}.}
\label{dis:tefflogg:tab}
\begin{center}
\begin{tabular}{lllll}
Name & T$_{\rm eff}$ (K) & $\log g$ & $\log (\rm He/H)$ & Model \\
\hline
KPD0025+5402  &28200  &5.37  &$-$2.9        &NLTE\\
PG0101+039    &27300  &5.50  &$-$2.7        &NLTE\\
PG0133+114    &29600  &5.66  &$-$2.3        &NLTE\\
PG0839+399    &37800  &5.53  &$-$3.7        &NLTE\\
PG1627+017    &21600  &5.12  &$-$2.9        &LTE\\
PG1716+426    &26100  &5.33  &$-$2.9        &LTE\\
PG1743+477    &27600  &5.58  &$-$1.8        &NLTE\\
HD171858      &27700  &5.25  &$-$2.9        &NLTE\\
KPD1946+4340  &34500  &5.37  &$-$1.35       &NLTE\\
\end{tabular}
\end{center}
\end{table}

\subsection{Orbital parameters known up to now}

Table~\ref{dis:tab} gives a list of all of the orbital parameters of
sdB binary stars known to date. In each case we require that both the
orbital period and the radial velocity semi-amplitude are measured. We
combine the orbital periods and the radial velocity semi-amplitudes to
calculate the mass function, $f_m$, of the system according to the
well-known relation:
\[f_m = \frac{M_2^3 \sin^3 i}{(M_1 +M_2)^2} = 
\frac{P K_{1}^3}{2 \pi G},\]
where the subscript ``$1$'' refers to the sdB star and ``$2$'' to its
companion.

If we take a canonical mass of $0.5\,\msun$ for the sdB star, we can
also calculate the minimum mass of its companion, $M_{2\mathrm{min}}$.
The values for $f_m$ and $M_{2\mathrm{min}}$ obtained in each case are
given in Table ~\ref{dis:tab}. We have also added in the table a
column indicating the nature of the companion, where known,
i.e. whether it is a compact object (most likely a white dwarf,
indicated by ``WD'') or a non-degenerate object (a main sequence star
or a brown dwarf, indicated by ``MS''). The evolution of each type is
fundamentally different since in the first case the system must go
through at least two mass transfer episodes whereas in the second case
the system suffers only one mass transfer episode. It is notable that
all the sdB stars with ``MS'' companions have very short orbital
periods.

\begin{table*}
\caption{List of all the sdBs with known orbital periods. The radial
  velocity semi-amplitude K, the minimum mass of the donor star M$_{2
  \rm min}$, the mass function f$_m$, and the companion type, either
  main sequence ``MS'' or white dwarf ``WD'',
  when known, are also given. A ``WD'' companion could
  actually be a white dwarf, a neutron star or another compact
  object. $^{1}$ indicates recent measurements looking for reflection
  effects in the lightcurves by Maxted et
  al. \shortcite{m02b}. 22 out of the 38 periods are measured in this
  paper. References for the other orbital periods given are (a) Moran
  et al. 1999, (b) Maxted et al. 2000a, (c) Dreschel et al. 2001, (d)
  Orosz \&\ Wade 1999, (e) Maxted et al. 2000b, (f) Wood \&\ Saffer
  1999, (g) Napiwotzki et al. 2001, (h) Saffer, Livio \&\ Yungelson 1998, (i)
  Kilkenny et al. 1998, (j) Edelmann, Heber \&\ Napiwotzki \shortcite{e01}, (k)
  Maxted et al. 2002a, (l) Foss, Wade \&\ Green 1991. $^*$ Edelmann,
  Heber \& Napiwotzki \shortcite{e01} measure values for this system consistent
  with the values presented in this paper.} 
\label{dis:tab}
\begin{center}
\begin{tabular}{lllllll}
Object&         P$_{\rm orb}$(d)&  K (km s$^{-1}$) &  M$_{2 \rm min}$
(M$_{\odot}$)& f$_m$ (M$_{\odot}$) & WD/MS & Ref.\\
\hline
PG0001+275&    0.528&         90.0& 0.293&  0.040&&  j \\
KPD0025+5402&  3.571&         40.2& 0.235&  0.024&&    \\
PG0101+039&    0.569908&     104.3& 0.370&  0.067&WD&   a \\
PG0133+114&    1.2382    &    83.2& 0.388&  0.074&&    $^*$\\
KPD0422+5421&  0.09017945&   237.0& 0.499&  0.124&WD&   d \\
HS0705+6700&   0.095646643&   85.8& 0.136&  0.006&MS&   c \\
PG0839+399&    5.622     &   33.5& 0.226 &  0.022&&   \\
PG0849+319&    0.7451    &    66.2& 0.228&  0.022&WD$^{1}$&    \\
PG0850+170&    27.81     &    33.5& 0.466&  0.108&&    \\
PG0907+123&    6.1163    &    59.8& 0.521&  0.136&&    \\
PG0918+029&    0.87679   &    79.9& 0.313&  0.046&WD$^{1}$&    \\
PG0940+068&    8.330     &    61.2& 0.634&  0.198&&   b \\
PG1017-086&    0.0729939 &    49.3& 0.066&  0.001&MS & k \\
PG1032+406&    6.779     &    33.7& 0.247&  0.027&&    \\
PG1043+760&    0.1201506 &    63.6& 0.106&  0.003&WD$^{1}$&    \\
HE1047-0436&   1.213253  &    94.0& 0.458&  0.104&WD&   g \\
PG1101+249&    0.35386   &   134.6& 0.424&  0.089&WD&   h \\
PG1110+294&    9.415     &    58.7& 0.633&  0.197&&   \\
PG1116+301&    0.85621   &    88.5& 0.356&  0.062&WD$^{1}$&    \\
HW~Vir&        0.116720  &    82.3& 0.140&  0.007&MS&   f \\
PG1247+554&    0.602740  &    32.2& 0.090&  0.002&&   b \\
PG1248+164&    0.73232   &    61.7& 0.207&  0.018&WD$^{1}$&    \\
PG1300+279&    2.2593    &    62.7& 0.346&  0.058&&    \\
PG1329+159&    0.249699  &    40.2& 0.083&  0.002&MS$^{1}$&    \\
PG1336-018&    0.1010174 &    78.0& 0.125&  0.005&MS&   i \\
PG1432+159&    0.22489   &   120.0& 0.294&  0.040&WD&   a \\
PG1512+244&    1.26978   &    92.7& 0.458&  0.105&&    \\
PG1538+269&    2.501     &    88.3& 0.600&  0.179&WD&   h, l \\
PG1619+522&    15.357    &    35.2& 0.376&  0.069&&    \\
PG1627+017&    0.829226  &    73.5& 0.273&  0.034&&    \\
PG1716+426&    1.77732   &    70.8& 0.366&  0.065&&    \\
PG1725+252&    0.601507  &   104.5& 0.381&  0.071&&    \\
UVO1735+22&    1.278     &   103.0& 0.539&  0.145&&  j \\
PG1743+477&    0.515561  &   121.3& 0.438&  0.095&&    \\
HD171858  &    1.529     &    93.6& 0.510&  0.130&&    \\
KPD1930+2752&  0.095111  &   349.3& 0.967&  0.420&WD&   e \\
KPD1946+4340&  0.403739  &   166.9& 0.628&  0.195&&    \\
PG2345+318  &  0.2409458 &   141.2& 0.379&  0.070&WD&   a \\
\end{tabular}
\end{center}
\end{table*}

\begin{figure}
\begin{picture}(100,0)(-270,250)
\put(0,0){\includegraphics{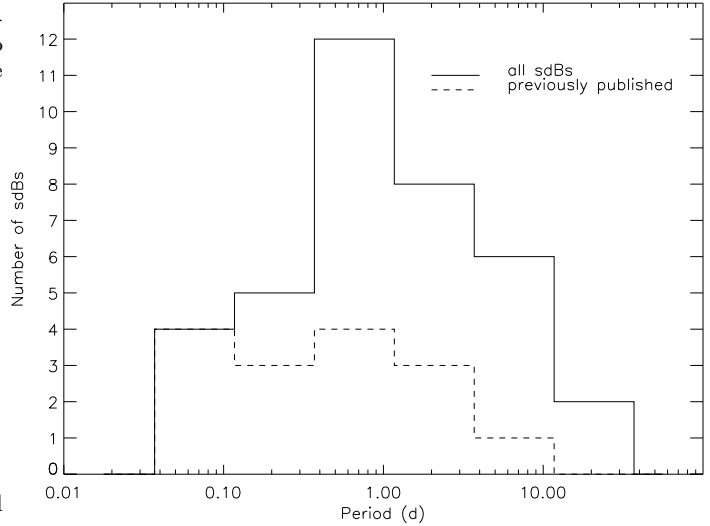}}
\noindent
\end{picture}
\vspace{70mm}
\caption{Histogram of orbital periods for all the sdB binaries known
  up to now.}
\label{dis:hist}
\end{figure}

In Fig.~\ref{dis:hist} we present a histogram of all the known orbital
periods of sdBs. The dashed line shows the systems with previously
published periods and the solid line shows the combination of the
previously published periods and the ones measured in this work.
There is no sign of any fine structure such as the well-known ``period
gap'' of the cataclysmic variable stars. The main points to take from
this plot are the large dynamic range and that our survey has
substantially increased the numbers of long period systems ($P >$ a
few days). This is probably a consequence of the large time base of
our data.

\begin{figure*}
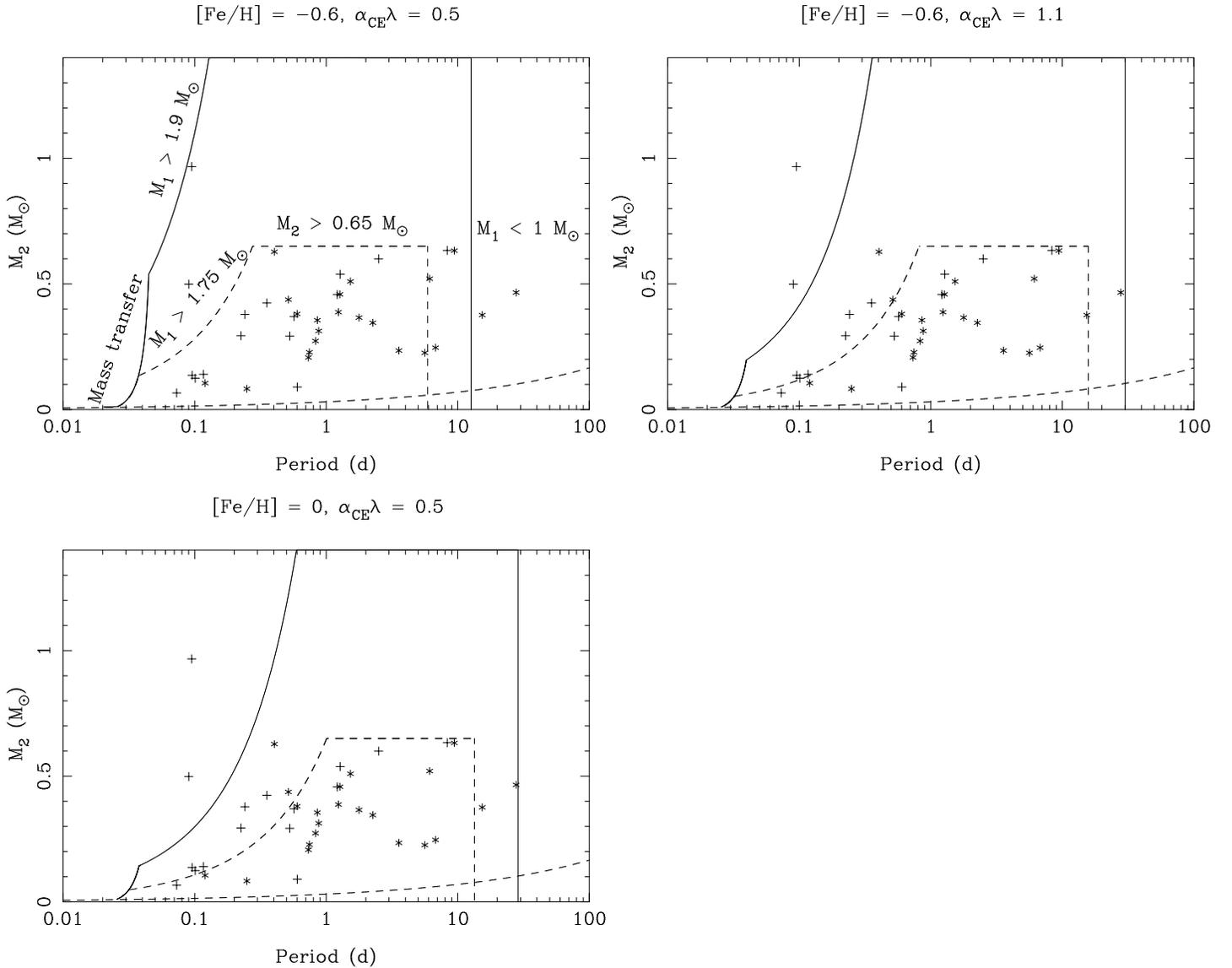

\begin{picture}(100,0)(10,20)
\put(0,0){\includegraphics{panel1.ps}}
\put(0,0){\includegraphics{panel2.ps}}
\put(0,0){\includegraphics{panel3.ps}}
\noindent
\end{picture}
\vspace{150mm}
\caption{Mass of the companion star versus orbital period. The sdB
  binaries whose periods are measured in this paper are represented by
  a * symbol whereas systems with previously published periods are
  represented by a +. We have also included in the plot different
  constraints. These are explained in detail in the text. The three
  panels present 3 different combinations of two parameters, the
  metallicity, [Fe/H], and the common envelope efficiency,
  $\alpha_{\rm CE}\lambda$.}
\label{dis:m2period1}
\end{figure*}

Fig.~\ref{dis:m2period1} shows $M_{2\mathrm{min}}$ versus orbital
period, assuming $M_1 = 0.5\,\msun$. All the systems given in
Table~\ref{dis:tab} are included in the plot. The systems whose
periods are measured in this paper are plotted as asterisks whereas
sdB binaries with previously published orbital periods are plotted as
plus signs. In addition, to test the idea of mass loss followed by a
common envelope shortly before the helium flash, we present limits
based upon a common envelope phase initiated when the progenitor of
the sdB star was at the tip of the red giant branch.  We calculate the
effect of the common envelope in the standard fashion \cite{w84} with
a fraction $\alpha_{\mathrm{CE}}$ of the loss of orbital energy set
equal to the binding energy of the envelope. We apply the standard
mass transfer equations instead of those calculated by Nelemans et al.
\shortcite{nypv01} because in most cases we are treating the second
phase of mass transfer instead of the first phase. The change in
orbital energy is
\[ \Delta E_{\mathrm{orb}} = \frac{G M_1 M_2}{2 a_i} -
\frac{G M_{\mathrm{sdB}} M_2}{2 a_f},\]
where $M_1$ is the mass of the
sdB star's progenitor and $a_i$ and $a_f$ are the initial and final
orbital separations. Parameterising the envelope binding energy as
\[ E_{\mathrm{env}} = - \frac{G M_1 (M_1 - M_{\mathrm{sdB}})}{\lambda R_1} ,\]
where $\lambda$ depends upon the structure of the envelope and the 
contribution of internal as well as gravitational energy 
\cite{dt00}. We then have
\[ \alpha_{\mathrm{CE}} \lambda \left(\frac{G M_{\mathrm{sdB}} M_2}{2 a_f} 
- \frac{G M_1 M_2}{2 a_i}\right) = 
\frac{G M_1 (M_1 - M_{\mathrm{sdB}})}{R_1} .\]
Thus given $M_1$, $M_2$, $M_{\mathrm{sdB}}$, $R_1$ and the combination
$\alpha_{\mathrm{CE}} \lambda$, the final separation can be computed
for any given initial separation.

Following this formalism, the different lines in the plots represent 
the following constraints:
\begin{enumerate}
\item the leftmost curve represents the limit imposed by the lack of
mass transfer in these binaries. The limit has been calculated
assuming that the companion is a main sequence star and therefore does
not apply for white dwarf companions. We have used Eggleton's
\shortcite{e83} formula for the Roche lobe radius to obtain this
limit.
  
\item the curve marked with $M_1 > 1.9\,\msun$ marks the limit beyond
  which there would be no helium flash. We have employed Hurley et
  al.'s \shortcite{h00} analytic formulae to calculate how large the
  progenitor of the sdB was, assuming that mass transfer occurred near
  the tip of the RGB. For a given initial mass, metallicity and common
  envelope efficiency one can then predict the current period as a
  function of $M_2$, assuming that the sdB star has a mass of
  $0.5\,\msun$ at the end. Massive companions get rid of the envelope
  easily and therefore end up at long periods. To get short periods
  and high companion masses one needs some combination of inefficient
  envelope ejection, a small red giant and a massive envelope. The
  latter two conditions occur for large mass progenitors since the
  radius at the tip of the giant branch decreases with mass for low
  mass stars. This sets a lower limit on the progenitor mass; $M_1 =
  1.9\,\msun$ is taken as the maximum since above this the core is not
  degenerate and there is then no natural explanation for the $\log
  g$/$T_{\rm eff}$ distribution of sdB stars.
  
\item the vertical line on the right hand side is the upper limit on
  the final period given that the sdB progenitor mass had to be $\ga
  1\,\msun$ for it to have evolved to this stage within the lifetime
  of the Galaxy.
  
\item the maximum value for $M_2$ has been chosen to be $1.4\,\msun$,
  corresponding to the Chandrasekhar stellar limit. Any main sequence
  star this massive would be easily visible.

\item the dashed line that appears at the bottom of the figures
indicates the minimum $M_2$ required to get the smallest amplitude
that we can detect ($\sim 10\,\mathrm{km}\,\mathrm{s}^{-1}$). This 
crudely shows that selection effects are likely to be weak in our data.

\end{enumerate}

In addition to show the sensitivity to variations of these limits, we
show the case for $M_1 < 1.75\,\msun$ and $M_2 < 0.65\,\msun$.  Most
of the data are contained within these more restrictive limits. It
should be remembered when looking at these plots that the companion
masses are lower limits owing to the unknown orbital inclinations of
most targets.

Three panels are shown to show the effect of varying either the
metallicity [Fe/H] (which affects the radius of the star) or the
common envelope efficiency/envelope structure parameter, $\alpha_{\rm
  CE}\lambda$. Increasing the metallicity increases the radius at mass
transfer and results in longer period systems being formed. This is
also the result if the efficiency with which orbital energy is used to
expel the common envelope increases. The largest value of $\alpha_{\rm
  CE}\lambda$ used in our calculations is 1.1. If the value of
$\alpha_{\rm CE}\lambda =$ 2, favoured by Nelemans et al.
\shortcite{nypv01}, is considered, this results in even longer period
systems being formed.

The conclusion to be drawn from all this is that although a common
envelope phase as a result of mass loss immediately prior to the
helium flash covers the range of parameters of most systems, it does
not explain simultaneously systems of long orbital period and systems
of short orbital period with massive companions. The two systems
KPD1930+2752 ($P = 0.0951\,\mathrm{d}$, $M_2 > 0.97\,\msun$) and
KPD0422+5421 ($P = 0.0902\,\mathrm{d}$, $M_2 > 0.50\,\msun$) look
particularly discrepant and may require an alternative formation path,
such as a descent from progenitors with $M_1 > 2\,\msun$.

\subsection{sdB stars with main-sequence companions}

Five of the short period sdB binary stars are known to have
main-sequence or brown dwarf companions. All of them have very short
periods indeed, with PG1329+159 having the longest period at
$4.0\,\mathrm{hr}$. This fits with the low masses of their companions,
$\sim 0.1\,\msun$ compared to $\sim 0.5\,\msun$ for a typical white
dwarf.  It also suggests an interesting possibility for the $10$ --
$20$ percent of sdB stars that are apparently single (Maxted et
al.\,2001; Green et al.\,2000): perhaps these stars could be the
result of merging of even lower mass companions, unable to survive the
common envelope phase. This would at least avoid the need to have a
single star route for sdB formation to supplement the binary star
route which already seem to account for 80 to 90 percent of these
stars. The long period sdB/main-sequence binaries identified by Green
et al.\ \shortcite{gls00} must have avoided a common envelope. Perhaps
they could do so because their relatively high mass main-sequence
components were able to accrete at a high enough rate to avoid the
common envelope. It would be of great interest to measure element
abundances in the main-sequence components of these stars, e.g.
abundance ratios of $^{12}$C/$^{13}$C \cite{sdmm95}.

\section{Conclusions}

We have confirmed the binary nature of 22 subdwarf B stars and have
measured their orbital parameters. This work increases the sample of
sdB binaries with known orbital parameters to 38. The observations
extend over several months allowing us to detect orbital periods of
the order of tens of days, longer than any previously measured.

We have measured T$_{\rm eff}$, $\log g$, $\log (\rm He/H)$ for the
sdBs where previous measurements of these quantities did not exist.
When we place the results in the T$_{\rm eff}$-$\log g$ plot we
find that two out of the 22 sdBs are post-EHB stars.

The large range of orbital periods and companion masses is a challenge
to simple theories for the formation of sdB stars. Although binary-induced
mass-loss at the tip of the red giant branch is able to explain most
systems, it appears unlikely to be the only formation route. Full
population synthesis will be needed to establish the viability of
alternative paths.

Amongst sdB stars with known orbits, those with low-mass main-sequence
or brown dwarf companions have particularly short periods. It seems
likely that a fraction of such systems, particularly those of very low
companion mass, may not have survived the common envelope phase. We
suggest that these could now be the single sdB stars.

\section*{Acknowledgements}

LMR was supported by a PPARC post-doctoral grant. The reduction and
analysis of the data were carried out on the Southampton node of the
STARLINK network. The Isaac Newton Telescope is operated on the island
of La Palma by the Isaac Newton Group in the Spanish Observatorio del
Roque de los Muchachos of the Instituto de Astrof\'{\i}sica de
Canarias. We thank PATT for their support of this program.


\begin{table*}
  \caption{Radial velocities measured for the 22 sdBs.}
  \label{results:rv:tab}
  \begin{center}
    \begin{tabular}{rrrrrrrr}

HJD & RV & HJD & RV & HJD & RV & HJD& RV\\
$-$2450000 & (km s$^{-1}$)& $-$2450000 & (km s$^{-1}$)& $-$2450000 &(km s$^{-1}$)& $-$2450000 &(km s$^{-1}$)\\
\hline
\multicolumn{2}{c}{\bf\underline{PG0133+114}} &
\multicolumn{2}{c}{\bf\underline{PG0849+319}} &
\multicolumn{2}{c}{\bf\underline{PG1032+406}} & \multicolumn{2}{c}{\bf
\underline{PG1116+301}} \\

2128.5749 & $-$72.7$\pm$3.5&2037.4053 &125.8$\pm$4.1&1649.3760 &$-$6.8$\pm$4.5&1653.5014 &$-$89.4$\pm$3.9\\
2128.5785 & $-$60.7$\pm$3.1& \multicolumn{2}{c}{\bf\underline{PG0850+170}} &1649.3776 &$-$15.8$\pm$3.6&1653.5164 &$-$86.4$\pm$2.8\\
2131.6335 & 76.2$\pm$1.8&1646.4208 &69.8$\pm$2.0&1651.4124 &24.5$\pm$2.5&1656.5285 &82.5$\pm$3.0\\
2131.6406 &  78.4$\pm$1.9&1646.4333 &67.3$\pm$2.2&1651.4141 &23.4$\pm$2.5&1656.5386 &79.3$\pm$2.8\\
2131.6824 &  79.6$\pm$2.5&1654.3966 &26.6$\pm$2.4&1946.5069 &13.0$\pm$4.5&1946.5774 &32.0$\pm$5.7\\
2131.6861 &  76.4$\pm$2.5&1654.4067 &25.1$\pm$2.3&1946.5088 &5.0$\pm$3.4&1946.5884 &39.0$\pm$4.9\\
2131.7340 &  96.2$\pm$2.9&1946.4409 &35.3$\pm$3.4&1978.5666 &61.9$\pm$1.5&1977.5184 &85.8$\pm$4.5 \\
2131.7376 &  90.5$\pm$2.7&1946.4518 &44.6$\pm$3.0&1978.5701 &58.4$\pm$1.5&1977.5288 &91.2$\pm$4.9\\
2133.6602 &  $-$78.3$\pm$2.0&1977.3953 &54.7$\pm$2.3&1979.6294 &41.4$\pm$1.3&1977.6297 &69.9$\pm$4.4\\
2133.6638 &  $-$77.6$\pm$2.0&1977.4057 &56.0$\pm$2.3&1979.6329 &40.8$\pm$1.3&1977.6401 &73.5$\pm$5.0\\
2133.7420 &  $-$55.0$\pm$2.1&1977.5829 &51.1$\pm$3.7&1979.7281 &38.2$\pm$2.4&1978.6363 &$-$7.2$\pm$3.1\\
2133.7456 & $-$54.8$\pm$2.1&1977.5933 &57.5$\pm$3.4&1979.7305 &32.2$\pm$2.6&1978.6468 &$-$7.9$\pm$3.9\\
2180.7593 &  $-$71.7$\pm$4.6&1978.3974 &67.3$\pm$3.5&1982.5671 &$-$2.7$\pm$2.0&1978.7442 &$-$56.9$\pm$11.7\\
2180.7629 &  $-$68.8$\pm$6.7&1978.4078 &68.1$\pm$2.9&1982.5695 &$-$4.4$\pm$1.9&1978.7547 &$-$62.1$\pm$14.7\\
2181.5733 & 12.0$\pm$2.3&1978.5355 &62.9$\pm$3.1&2032.3651 &56.7$\pm$2.2&1982.5794 &59.4$\pm$2.7\\
2181.5769 & 9.6$\pm$3.3&1979.4663 &68.1$\pm$2.2&2032.3698 &54.3$\pm$2.4&1982.5898 &64.8$\pm$2.8\\
2187.7392 & 22.9$\pm$1.9&1979.4767 &62.9$\pm$2.3&2033.3479 &56.2$\pm$3.2&\multicolumn{2}{c}{\bf \underline{PG1248+164}}\\
2187.7428 & 22.1$\pm$1.9&1979.5629 &67.5$\pm$3.2&2033.3503 &55.5$\pm$2.5&1646.5672 &41.5$\pm$2.0\\
\multicolumn{2}{c}{\bf \underline{PG0839+399}}&1979.5681 &61.2$\pm$3.5&2033.3536 &50.1$\pm$2.4&1646.5875 &45.5$\pm$1.9\\
1646.3498 &53.5$\pm$8.3&1982.4753 &64.8$\pm$2.6&2033.3560 &49.7$\pm$2.3&1656.5529 &$-$63.6$\pm$3.2\\
1646.3562 &32.4$\pm$8.4&1982.4857 &59.2$\pm$2.4&2035.5349 &1.5$\pm$1.1&1656.5632 &$-$69.2$\pm$3.2\\
1651.3528 &67.9$\pm$4.5&2032.4352 &60.7$\pm$2.1&2035.5419 &$-$7.8$\pm$1.1&1946.6313 &$-$46.2$\pm$4.9\\
1651.3677 &55.9$\pm$4.2&2032.4560 &51.7$\pm$1.8&2036.3546 &$-$7.8$\pm$4.3&1946.6422 &$-$22.4$\pm$7.5\\
1654.3695 &$-$7.0$\pm$4.9&2037.4162 &62.0$\pm$2.2&2036.3570 &$-$7.8$\pm$3.7&1977.6948 &28.9$\pm$6.3\\
1654.3820 &$-$8.2$\pm$4.6&2037.4266 &61.2$\pm$2.1&\multicolumn{2}{c}{\bf \underline{PG1043+760}} &1977.7052 &23.5$\pm$5.9\\
1946.3902 &$-$11.3$\pm$4.2&\multicolumn{2}{c}{\bf \underline{PG0907+123}} &1649.3849 &$-$27.1$\pm$3.3&1982.7060 &44.0$\pm$4.6\\
1977.3520 &55.1$\pm$3.6&1647.4116 &86.4$\pm$2.9&1649.3959 &$-$4.0$\pm$3.4&1982.7164 &52.0$\pm$4.4\\
1977.4229 &52.0$\pm$3.5&1647.4323 &84.1$\pm$2.5&1651.4332 &$-$21.9$\pm$3.3&2036.5834 &$-$76.1$\pm$3.2\\
1977.6104 &54.4$\pm$4.4&1653.3980 &98.6$\pm$4.3&1651.4402 &$-$4.8$\pm$3.2&2036.5939 &$-$77.3$\pm$3.2 \\
1978.3634 &37.1$\pm$4.7&1653.4084 &94.1$\pm$7.2&1978.5149 &77.4$\pm$4.2&2037.5674 &28.5$\pm$4.2\\
1978.4527 &33.7$\pm$5.0&1654.4213 &34.4$\pm$3.7&1978.5219 &84.2$\pm$3.7&2037.5779 &26.7$\pm$4.5\\
1979.3939 &$-$1.2$\pm$4.6&1654.4351 &37.9$\pm$2.1&1979.4291 &$-$41.6$\pm$5.7&2038.5088 &16.4$\pm$6.1\\
1979.4044 &0.3$\pm$4.6&1946.4705 &111.4$\pm$2.3&1979.4360 &$-$51.5$\pm$5.3&2038.5193 &13.3$\pm$4.6\\
1982.4504 &62.9$\pm$7.1&1946.4849 &114.4$\pm$2.7&1979.6126 &85.7$\pm$2.6&\multicolumn{2}{c}{\bf \underline{PG1300+279}}\\
1982.4608 &48.1$\pm$6.4&1978.4914 &37.9$\pm$2.6&1979.6195 &81.2$\pm$2.6&1646.6051 &53.0$\pm$2.4\\
2033.3690 &39.8$\pm$7.6&1978.5053 &36.6$\pm$2.5&1979.7118 &63.0$\pm$3.6&1646.6166 &50.2$\pm$2.3\\
2033.3795 &61.1$\pm$6.8&1979.4910 &$-$1.8$\pm$1.8&1979.7188 &78.3$\pm$3.3&1656.6042 &$-$33.9$\pm$2.6\\
2037.3763 &7.9$\pm$5.3&1979.5049 &0.5$\pm$1.8&2035.5512 &$-$21.8$\pm$2.8&1656.6156 &$-$41.9$\pm$2.5\\
2037.3868 &11.4$\pm$5.2&1982.3503 &105.4$\pm$3.5&2035.5581 &$-$14.4$\pm$2.6&1946.6580 &$-$15.3$\pm$5.0\\
2038.3738 &56.0$\pm$10.0&1982.3642 &107.8$\pm$3.7& \multicolumn{2}{c}{\bf \underline{PG1110+294}}&1946.6690 &$-$10.1$\pm$4.0\\
2038.4482 &43.1$\pm$7.4&2032.3927 &104.7$\pm$2.2&1653.4747 &8.7$\pm$13.1&1977.7172 &$-$69.0$\pm$4.9\\
2181.7460 &20.2$\pm$8.8&2032.4136 &112.6$\pm$3.3&1653.4845 &9.6$\pm$10.3&1977.7276 &$-$66.2$\pm$3.7\\
2181.7602 &$-$4.2$\pm$5.7& \multicolumn{2}{c}{\bf \underline{PG0918+029}} &1656.5069 &1.2$\pm$2.9&1979.6646 &$-$33.3$\pm$2.6\\
\multicolumn{2}{c}{\bf \underline{PG0849+319}}&1647.4808 &35.3$\pm$5.9&1656.5167 &1.3$\pm$2.9&1979.6750 &$-$36.0$\pm$2.8\\
1646.3657& 92.2$\pm$4.3&1647.4861 &26.4$\pm$11.1&1657.4808 &$-$39.8$\pm$3.1&1982.6567 &$-$37.4$\pm$3.1\\
1646.3733 &82.2$\pm$4.2&1653.3633 &96.2$\pm$5.0&1657.4906 &$-$40.1$\pm$3.1&1982.6672 &$-$33.9$\pm$3.3\\
1651.3816 &121.2 $\pm$4.1&1653.3669 &107.5$\pm$4.5&1946.5472 &28.6$\pm$5.5&1982.7663 &$-$23.4$\pm$4.4\\
1651.3891 &128.7$\pm$4.1&1977.4392 &148.6$\pm$5.1&1946.5633 &40.8$\pm$4.6&1982.7732 &$-$20.8$\pm$4.5\\
1946.4116 &103.4$\pm$4.6&1977.4428 &159.4$\pm$4.5&1978.6659 &$-$64.0$\pm$4.5&2036.6295 &$-$58.2$\pm$2.9\\
1946.4191 &122.8$\pm$6.5&1978.3468 &185.9$\pm$5.5&1978.6764 &$-$66.6$\pm$4.3&2036.6405 &$-$66.6$\pm$2.8\\
1946.4268 &121.5$\pm$6.8&1978.3503 &163.6$\pm$5.0&1978.7165 &$-$72.0$\pm$5.9&\multicolumn{2}{c}{\bf \underline{PG1329+159}} \\
1977.3726 &7.0$\pm$5.3&1978.4363 &178.6$\pm$5.4&1978.7304 &$-$52.9$\pm$6.8&1646.6683 &$-$1.2$\pm$3.1\\
1977.3798 &2.7$\pm$5.4&1978.4398 &186.9$\pm$5.4&1979.6421 &$-$74.8$\pm$2.7&1646.6720 &$-$9.9$\pm$3.0\\
1978.3795 &81.5$\pm$7.7&1978.5994 &132.8$\pm$4.8&1979.6526 &$-$74.9$\pm$2.8&1656.6277 &17.0$\pm$2.2\\
1978.3865 &86.6$\pm$6.1&1978.6029 &135.1$\pm$4.0&1982.5328 &18.6$\pm$2.7&1656.6351 &10.9$\pm$2.2\\
1978.4704 &112.1$\pm$5.5&1979.5487&90.5$\pm$2.5&1982.5433 &17.4$\pm$2.8&1947.7337 &1.2$\pm$5.6\\
1978.4774 &142.9$\pm$5.2&1979.5556 &88.7$\pm$2.7&1982.6798 &20.1$\pm$3.8&1947.7411 &3.7$\pm$7.4\\
1979.4152 &85.8$\pm$3.3&2036.3670&198.5$\pm$9.2&1982.6903 &24.1$\pm$4.4&1977.6529 &$-$31.8$\pm$3.1\\
1979.4221 &81.4$\pm$3.8&2036.3740 &173.7$\pm$8.9&2032.5190 &19.0$\pm$2.4&1977.6599 &$-$25.9$\pm$3.2\\
1982.3773 &97.7$\pm$11.8&2037.4384 &82.2$\pm$2.4&2032.5346 &25.0$\pm$3.4&1978.6865 &$-$6.0$\pm$2.9\\
1982.4969 &33.0$\pm$4.3&2037.4453 &75.1$\pm$2.6&2034.5643 &$-$38.9$\pm$10.8&1978.6934 &4.2$\pm$3.2\\
1982.5039 &24.6$\pm$4.4&&& &&1978.7651 &8.3$\pm$7.2\\
2037.3983 &139.1$\pm$4.4&&& &&1978.7721 &3.9$\pm$7.1\\
   \end{tabular}
  \end{center}
\end{table*}
\setcounter{table}{5}
\begin{table*}
  \caption{continued.}
  \begin{center}
    \begin{tabular}{rrrrrrrr}
HJD & RV &HJD & RV & HJD & RV & HJD & RV \\
$-$2450000 & (km s$^{-1}$) & $-$2450000 & (km s$^{-1}$)& $-$2450000 & (km s$^{-1}$)& $-$2450000 & (km s$^{-1}$)\\
\hline

\multicolumn{2}{c}{\bf \underline{PG1329+159}} &\multicolumn{2}{c}{\bf \underline{PG1627+017}}&\multicolumn{2}{c}{\bf \underline{PG1725+252}} &\multicolumn{2}{c}{\bf \underline{HD171858}}\\
1979.6859 &$-$0.3$\pm$2.6&1653.6236 &$-$125.4$\pm$3.6&1651.7613 &$-$62.2$\pm$6.7&2132.4162 &135.8$\pm$0.8\\
1979.6929 &4.3$\pm$2.2&1653.6261 &$-$126.9$\pm$3.5&1651.7629 &$-$63.7$\pm$9.5&2132.4186 &136.5$\pm$0.8\\
1982.6163 &$-$57.4$\pm$2.4&1654.6144 &$-$66.6$\pm$2.7&1653.7578 &40.1$\pm$3.7&2132.4211 &136.1$\pm$0.8\\
1982.6233 &$-$52.3$\pm$2.3&1654.6169 &$-$68.0$\pm$2.8&1653.7610 &39.7$\pm$4.7&2132.4974 &153.5$\pm$0.9\\
2033.4229 &12.6$\pm$2.9&1978.7757 &$-$113.8$\pm$6.2&1654.7494 &$-$93.5$\pm$1.8&2132.4998 &154.8$\pm$0.8\\
2033.4298 &13.6$\pm$2.7&1978.7792 &$-$98.9$\pm$6.8&1654.7554 &$-$86.7$\pm$2.0&2132.5023 &158.2$\pm$0.8\\
2033.6309 &14.9$\pm$5.1&2032.7251 &$-$71.9$\pm$1.7&1977.7686 &$-$78.2$\pm$3.3&2133.3961 &$-$21.9$\pm$0.9\\
2033.6379 &15.5$\pm$8.9&2032.7321 &$-$66.7$\pm$1.7&1977.7721 &$-$77.3$\pm$3.2&2133.3985 &$-$17.8$\pm$0.9\\
2033.6493 &14.6$\pm$4.2&2032.7397 &$-$61.4$\pm$1.7&1979.7769 &36.7$\pm$2.2&2133.4728 &$-$12.3$\pm$1.3\\
2034.6203 &4.3$\pm$2.0 &2032.7464 &$-$59.1$\pm$2.3&1979.7804 &32.8$\pm$2.3&2133.4752 &$-$13.0$\pm$1.3\\
2034.6342 &14.8$\pm$1.7&2033.5451 &$-$77.7$\pm$2.2&1979.7839 &34.2$\pm$2.4&2133.5059 &$-$10.9$\pm$0.8\\
\multicolumn{2}{c}{\bf \underline{PG1512+244}} &2033.5486 &$-$72.2$\pm$2.4&1979.7880 &34.9$\pm$2.9&2133.5083 &$-$7.6$\pm$0.8\\
1653.5782 &$-$101.6$\pm$3.2&2033.6187 &$-$35.5$\pm$2.2&1982.7842 &36.8$\pm$3.5&\multicolumn{2}{c}{\bf \underline{KPD0025+5402}}\\
1653.5837 &$-$94.7$\pm$3.0&2033.6222 &$-$36.1$\pm$2.2&1982.7865 &39.9$\pm$3.6&2129.6652 &$-$38.2$\pm$2.2\\
1657.6469 &$-$42.9$\pm$2.8&2033.6614 &$-$18.3$\pm$3.5&1982.7888 &22.9$\pm$4.7&2129.6792 &$-$40.0$\pm$2.6\\
1657.6525 &$-$41.0$\pm$2.8&2033.6650 &$-$10.7$\pm$3.3&2032.6991 &42.5$\pm$1.8&2130.7039 &$-$9.4$\pm$2.2\\
1946.6963 &$-$46.0$\pm$3.2&2035.5739 &9.5$\pm$2.1&2032.7061 &42.4$\pm$1.7&2130.7180 &$-$7.5$\pm$2.2\\
1946.7003 &$-$40.3$\pm$3.2&2035.5774 &2.0$\pm$2.1&2032.7130 &39.9$\pm$1.7&2132.5951 &$-$4.7$\pm$2.5\\
1978.6991 &$-$90.4$\pm$4.1&2035.5814 &0.0$\pm$2.2&2033.5694 &$-$164.0$\pm$2.4&2132.6057 &0.0$\pm$2.5\\
1978.7026 &$-$97.0$\pm$4.0&2035.5849 &$-$4.0$\pm$2.6&2033.5729 &$-$159.9$\pm$2.5&2132.7141 &$-$14.5$\pm$3.1\\
1979.7629 &$-$56.4$\pm$2.9&2036.7411 &$-$118.5$\pm$1.6&2033.5764 &$-$165.4$\pm$2.4&2132.7212 &$-$18.7$\pm$3.0\\
1979.7664 &$-$62.8$\pm$2.9&2036.7463 &$-$115.0$\pm$2.7&2033.7145 &$-$76.7$\pm$3.1&2133.4828 &$-$52.6$\pm$3.1\\
1979.7708 &$-$64.6$\pm$2.8&2037.7194 &$-$62.7$\pm$1.8&2033.7180 &$-$71.9$\pm$3.2&2133.4933 &$-$56.8$\pm$3.2\\
1979.7743 &$-$59.9$\pm$2.9&2037.7264 &$-$57.9$\pm$1.8&2033.7215 &$-$73.6$\pm$3.1&2133.5459 &$-$42.0$\pm$2.6\\
1982.7785 &$-$13.9$\pm$3.4&2037.7341 &$-$50.8$\pm$1.7&2033.7438 &$-$54.3$\pm$5.5&2133.5565 &$-$44.4$\pm$2.6\\
1982.7820 &$-$9.0$\pm$3.7&2037.7388 &$-$53.5$\pm$3.6&2033.7473 &$-$39.3$\pm$6.1&2133.6730 &$-$50.5$\pm$2.2\\
2032.5857 &85.9$\pm$2.3&2128.4616 &12.4$\pm$3.1&2034.6777 &$-$107.5$\pm$1.9&2133.6835 &$-$52.0$\pm$2.2\\
2033.5114 &$-$35.4$\pm$3.4&2128.4652 &8.5$\pm$3.0&2034.6829 &$-$119.7$\pm$3.3&2182.6964 &$-$18.9$\pm$1.9\\
2033.5149 &$-$35.6$\pm$3.9&2129.3915 &$-$35.4$\pm$6.3&2128.4364 &$-$31.7$\pm$4.8&2182.7070 &$-$17.7$\pm$1.8\\
2033.5193 &$-$38.7$\pm$3.7&2129.3951 &$-$39.4$\pm$3.5&2128.4400 &$-$37.7$\pm$4.9&2183.7615 &$-$45.2$\pm$2.3\\
2129.3764 &42.3$\pm$6.5&2183.3209 &$-$59.3$\pm$2.8&\multicolumn{2}{c}{\bf \underline{PG1743+477}} &2183.7686 &$-$41.7$\pm$4.6\\
2129.3800 &7.6$\pm$10.5&2183.3245 &$-$60.0$\pm$2.6&1653.7123 &38.9$\pm$2.2&2184.6462 &4.9$\pm$2.1\\
 \multicolumn{2}{c}{\bf \underline{PG1619+522}}& \multicolumn{2}{c}{\bf \underline{PG1716+426}}&1653.7220 &28.9$\pm$2.3&2184.6567 &7.3$\pm$3.2\\
1646.7032 &$-$69.3$\pm$3.7&1646.7552 &$-$53.1$\pm$2.4&1655.6932 &39.8$\pm$1.9&2187.7147 &$-$28.1$\pm$2.2\\
1646.7081 &$-$76.0$\pm$3.6&1651.5972 &59.6$\pm$2.5&1655.7029 &45.9$\pm$1.9&2187.7253 &$-$21.7$\pm$2.2\\
1651.5444 &$-$67.1$\pm$3.5&1651.6078 &49.2$\pm$2.6&1656.7376 &45.7$\pm$2.2&\multicolumn{2}{c}{\bf \underline{KPD1946+4340}}\\
1651.5518 &$-$68.1$\pm$3.7&1657.7333 &$-$73.2$\pm$2.1&1656.7474 &50.1$\pm$2.3&2130.4461 &104.4$\pm$2.8\\
1653.6887 &$-$35.0$\pm$4.3&1657.7481 &$-$71.3$\pm$2.1&1946.7770 &$-$183.7$\pm$2.7&2130.4601 &130.0$\pm$2.9\\
1653.6937 &$-$40.4$\pm$4.2&1946.7530 &31.9$\pm$3.1&1946.7880 &$-$183.7$\pm$2.6&2131.4134 &$-$13.6$\pm$3.0\\
1946.7090 &$-$33.4$\pm$3.3&1946.7640 &30.8$\pm$3.5&2032.6488 &39.6$\pm$1.4&2131.4274 &$-$46.2$\pm$2.8\\
1946.7164 &$-$29.2$\pm$3.2&1977.7568 &$-$59.3$\pm$2.1&2032.6645 &32.0$\pm$2.0&2131.4652 &$-$119.9$\pm$4.5\\
1977.7384 &$-$20.9$\pm$2.2&1979.7514 &$-$12.3$\pm$2.2&2033.4885 &$-$100.8$\pm$3.0&2131.4723 &$-$137.1$\pm$4.5\\
1979.7005 &$-$16.5$\pm$2.5&2032.6807 &$-$71.9$\pm$2.1&2033.4989 &$-$88.3$\pm$2.7&2131.4794 &$-$147.2$\pm$4.5\\
2033.5868 &$-$89.0$\pm$2.3&2033.5284 &67.1$\pm$2.3&2035.6308 &17.4$\pm$2.1&2131.4865 &$-$158.8$\pm$4.4\\
2033.7010 &$-$90.6$\pm$2.5&2184.3237 &24.0$\pm$4.3&2035.6412 &28.9$\pm$2.0&2131.5994 &$-$44.7$\pm$4.8\\
2034.6621 &$-$75.5$\pm$1.9&2184.3273 &31.9$\pm$4.3&2035.6894 &53.9$\pm$1.6&2131.6065 &$-$28.6$\pm$4.8\\
2035.5945 &$-$65.9$\pm$1.9&&&2035.6998 &59.1$\pm$1.7&2184.4898 &$-$43.5$\pm$3.4\\
&&&&2188.3633 &27.4$\pm$2.8&2184.4969 &$-$24.8$\pm$3.5\\
&&&&2188.3669 &30.5$\pm$2.8&2187.5088 &49.8$\pm$3.7\\
&&&&&&2187.5159 &26.5$\pm$3.7\\
   \end{tabular}
  \end{center}
\end{table*}

\appendix
 
\section{The probability of being wrong}
 
Marsh, Dhillon \&\ Duck \shortcite{mdd95} derived the following
equation (their A4) for the probability of a binary star of orbital
frequency $f$ versus a single star given a set of radial velocity
data, $D$
\begin{eqnarray*}
\frac{P(B,f|D)}{P(S|D)}
&=& \frac{P(B,f)}{P(S)} \frac{4 \left(2\pi\right)^{3/2}}{R_K^2 R_\gamma \left(\det\mathbf{A}\right)^{1/2}}
\frac{ R_\gamma \left(\sum_i w_i \right)^{1/2}}{\left(2\pi\right)^{1/2}} \nonumber \\
& & \times \exp \frac{1}{2} \left( \mathbf{b}^t \mathbf{A}^{-1} \mathbf{b} - \frac{\left(\sum_i w_i V_i \right)^2}{\sum_i w_i}
\right), \label{eq:integral}
\end{eqnarray*}
where the vector $\mathbf{b}$ is given by
\begin{equation}
\mathbf{b} = \left( \begin{array}{c} \sum_i w_i \\ \sum_i w_i c_i \\ \sum_i w_i
s_i \end{array} \right),
\end{equation}
and the matrix $\mathbf{A}$ is
\begin{equation}
\mathbf{A} = \left(
\begin{array}{ccc}
\sum_i w_i & \sum_i w_i c_i & \sum_i w_i s_i \\
\sum_i w_i c_i & \sum_i w_i c^2_i & \sum_i w_i c_i s_i \\
\sum_i w_i s_i& \sum_i w_i s_i c_i & \sum_i w_i s^2_i
\end{array}
\right),
\end{equation}

with $w_i = 1/\sigma^2_i$, where $\sigma_i$ is the uncertainty on the
$i$-th point and $c_i = \cos(2\pi f t_i)$, $s_i = \sin(2\pi f t_i)$
where $t_i$ is the time of the $i$-th point; $V_i$ is the measured
radial velocity for the $i$-th point. The term inside the brackets of
the exponential is equal to the difference in $\chi^2$ between
adopting a constant velocity model versus a model of a constant plus a
sinusoid, $\chi^2_c - \chi^2_s$.  That is, it is a measure of the gain
one makes by adding a sinusoid to a constant model.  Since the
constant model $\chi^2_c$ is independent of frequency, this is the
justification for our statement in the main text that the probability
of a certain period is dominated by the term $\exp - \chi^2/2$.

Already some assumptions have gone into this result which is the end
product of integrating over the systemic velocity and the sine and
cosine amplitudes for a given frequency. In particular we took the
prior probabilities over these parameters to be uniform in a box
extending from $-R_\gamma$ to $+R_\gamma$ in systemic velocity, and
$-R_K$ to $+R_K$ in each of the semi-amplitudes (a factor 4 has
entered into Eq.~\ref{eq:integral} compared to the original version
from Marsh, Dhillon \&\ Duck \shortcite{mdd95} to account for a factor
2 change in our definition of these ranges). We continue to adopt this
prior, largely because it then leads to a tractable integral, but also
because it is hard to justify any particular choice of prior, and so
one might as well adopt the simplest (although see below for more on
this).

Eq.~\ref{eq:integral} is correct, given our assumptions, as long as
the volume of integration wholly encloses the integrand in the
systemic velocity/semi-amplitude space. The integrand is a 3D Gaussian
centred upon the best-fit values. For good enough data, this will
occupy a small region, comparable in dimension to the uncertainties
quoted on these parameters. At least this is the case for the best-fit
frequency, but it is less clear that it will always be the case. To
account for the case where a significant fraction of the integrand
lies outside the integration region, we wrote a program that
calculated the value of Eq.~\ref{eq:integral} and, in addition, a
correction to it from a Monte Carlo integration. Essentially this
computes the fraction of the integrand lying outside the volume of
integration.  A detail was that we changed the volume of integration
to a cylinder with axis along the systemic velocity axis and radius
$R_K$, which gives a uniform prior in orbital phase as one expects.
The Monte Carlo work was carried out by computing the eigenvalues and
eigenvectors of the matrix $\mathbf{A}$ and then generating points
scattered around the maximum using Gaussian random number generators
to define the multiple of each eigenvector to use. The correction
factor was then the ratio of the number of points that fell within the
volume of integration to the total number of points generated. In
practice our data was good enough that none of this made much
difference to the results.

We adopted a value of $R_\gamma = 500\,\mathrm{km}\,\mathrm{s}^{-1}$
and set $R_K$ by taking the maximum companion mass to be
$1.5\,\mathrm{M}_\odot$. There is certainly a case for having a prior
in the systemic velocity that falls off more gradually from a peak
close to zero. A Gaussian for example could still be accommodated
analytically by a small modification of the matrix $\mathbf{A}$,
however we feel that the greater ease of understanding of a uniform
prior is an advantage that should not be wasted. What this means in
practice is that we will over- or under-estimate the probabilities
listed in Table~\ref{tab:probs} according to whether the best-fit
systemic velocity is far from or close to zero velocity. This is a
quantitative manifestation of the lack of confidence one would have in
an orbit with a systemic velocity of, say,
$2000\,\mathrm{km}\,\mathrm{s}^{-1}$ compared to a more moderate one.
Similar remarks apply to the prior for $K$. The uncertainty over the
priors is why the final probabilities are themselves subject to
uncertainty.

The final step was to integrate the probability as a function of
frequency over ranges of frequency. The only technical difficulty here
is that the integrand over frequency is extremely peaky with large
ranges of very low probability pierced by narrow spikes of high
probability. There is little one can do about this apart from
computing the integrand over a very finely spaced series of
frequencies. We applied straightforward trapezoidal integration,
halving the spacing of the frequency grid until the integral
converged. Typically this required of the order of one million points,
but did not take overly long on any one object. The same points about
the systemic velocity and semi-amplitude priors applies to the prior
over frequency. We took a prior uniform in frequency, but this has no
particular justification. The final probability is uncertain to the
extent that this prior deviates from reality. A factor of 10 would not
especially surprise us here.

\label{lastpage}


\begin{thebibliography}{1}

\bibitem[\protect\citename{Brassard et al.\,}2001]{b01} Brassard P.,
Fontaine G., Billeres M., Charpinet S., Liebert J.,Saffer R.\,A.,
2001, ApJ, 563, 1013

\bibitem[\protect\citename{Cumming, Marcy \&\ Butler\,}1999]{cmb99}
  Cumming\,A., Marcy\,G.\,W., Butler\,R.\,P., 1999, ApJ, 526, 890
  
\bibitem[\protect\citename{D'Cruz et al.\,}1996]{dc96} D'Cruz\,N.\,L.,
  Dorman\,B., Rood\,R.\,T., O'Connell\,R.\,W., 1996, ApJ, 466, 359

\bibitem[\protect\citename{Drechsel et al.\,}2001]{d01} Drechsel\,H.,
  Heber\,U., Napiwotzki\,R., Ostensen\,R., Solheim\,J-E.,
  Johannessen\,F., Schuh\,S.\,L., Deetjen\,J., Zola\,S., 2001, A\&A,
  379, 893
  
\bibitem[\protect\citename{Dewi \&\ Tauris\,}2000]{dt00} Dewi\,J.\,D.\,M.,
  Tauris\,T.\,M., 2000, A\&A, 360, 1043

\bibitem[\protect\citename{Edelmann, Heber \&\ Napiwotzki\,}2001]{e01}
  Edelmann\,H., Heber\,U., Napiwotzki\,R., 2001, AN, 322, 401

\bibitem[\protect\citename{Eggleton\,}1983]{e83} Eggleton\,P.\,P., 1983,
  ApJ, 268, 368

\bibitem[\protect\citename{Foss, Wade \&\ Green\,}1991]{f91} Foss\,D.,
Wade\,R.\,A., Green\,R.\,F., 1991, ApJ, 374, 281

\bibitem[\protect\citename{Green, Liebert \&\ Saffer\,}2000]{gls00}
  Green\,E.\,M., Liebert\,J., Saffer\,R.\,A., 2000, Proceedings of
  the Twelfth European Workshop on White Dwarfs. ASP Conference
  Series, in press (astro-ph/0012246)

\bibitem[\protect\citename{Heber et al.\,}1984]{h84} Heber\,U.,
  Hunger\,K., Jonas\,G., Kudritzki\,R.\,P., 1984, A\&A, 130, 119

\bibitem[\protect\citename{Heber, Reid \&\ Werner\,}2000]{hrw00}
  Heber\,U., Reid\,I.\,N., Werner\,K., 2000, A\&A, 363, 198


\bibitem[\protect\citename{Hurley et al.\,}2000]{h00} Hurley\,J.\,R.,
   Pols\,O.\,R., Tout\,C.\,A., 2000, MNRAS, 315, 543 

\bibitem[\protect\citename{Kilkenny et al.\,}1998]{k98} Kilkenny\,D.,
  O'Donoghue\,D., Koen\,C., Lynas-Gray\,A.\,E., van Wyk\,F., 1998,
  MNRAS, 296, 329

\bibitem[\protect\citename{Lomb\,}1976]{l76} Lomb\,N.\,R., 1976, Ap\&SS,39,447

\bibitem[\protect\citename{Marsh\,}1989]{m89} Marsh\,T.\,R., 1989,
PASP, 101, 1032

\bibitem[\protect\citename{Marsh, Dhillon \&\ Duck\,}1995]{mdd95}
  Marsh\,T.\,R., Dhillon\,V.\,S., Duck\,S.\,R., 1995, MNRAS, 275, 828

\bibitem[\protect\citename{Maxted et al.\,}2001]{m01}
  Maxted\,P.\,F.\,L., Heber\,U., Marsh\,T.\,R., North\,R.\,C., 2001,
  MNRAS, 326, 139
  
\bibitem[\protect\citename{Maxted et al.\,}2002b]{m02b}
  Maxted\,P.\,F.\,L., Marsh\,T.\,R., Morales-Rueda\,L., MNRAS, 2002b, in
  preparation
  
\bibitem[\protect\citename{Maxted et al.\,}2002a]{m02a}
  Maxted\,P.\,F.\,L., Marsh\,T.\,R., Heber\,U., Morales-Rueda\,L.,
  North R.\,C., Lawson\,W.\,A., 2002a, MNRAS, 333, 231

\bibitem[\protect\citename{Maxted et al.\,}2000c]{m00c}
  Maxted\,P.\,F.\,L., Marsh\,T.\,R., Moran\,C.\,K.\,J., 2000c, MNRAS, 319, 305 

\bibitem[\protect\citename{Maxted et al.\,}2000b]{m00b}
  Maxted\,P.\,F.\,L., Marsh\,T.\,R., North\,R.\,C., 2000b, MNRAS, 317, 41

\bibitem[\protect\citename{Maxted et al.\,}2000a]{m00a}
  Maxted\,P.\,F.\,L., Moran\,C.\,K.\,J., Marsh\,T.\,R., Gatti\,A.\,A.,
  2000a, MNRAS, 311, 877

\bibitem[\protect\citename{Moran et al.\,}1999]{m99} Moran\,C.,
  Maxted\,P., Marsh\,T.\,R., Saffer\,R.\,A., Livio\,M., 1999, MNRAS,
  304, 535
  
\bibitem[\protect\citename{Napiwoztki\,}1997]{n97} Napiwotzki\,R.,
  1997, A\&A, 322, 256

\bibitem[\protect\citename{Napiwotzki et al.\,}2001]{n01} Napiwotzki\,R.,
  Edelmann\,H., Heber\,U.,Karl\,C., Drechsel\,H., Pauli\,E-M.,
  Christlieb\,N., 2001, A\&A, 378, 17
  
\bibitem[\protect\citename{Nelemans et al.\,}2001]{nypv01}
  Nelemans\,G., Yungelson\,L.\,R., Portegies Zwart\,S.\,P.,
  Verbunt\,F., 2001, A\&A, 365, 491

\bibitem[\protect\citename{Orosz \&\ Wade\,}{1999}]{ow99}
  Orosz\,J.\,A., Wade\,R.\,A., 1999, MNRAS, 310, 773

\bibitem[\protect\citename{Saffer et al.\,}1994]{s94} Saffer\,R.\,A.,
  Bergeron\,P., Koester\,D., Liebert\,J., 1994, ApJ, 432, 351


\bibitem[\protect\citename{Saffer et al.\,}1998]{s98} Saffer\,R.\,A.,
  Livio\,M., Yungelson\,L.\,R., 1998, ApJ, 502, 394
  
\bibitem[\protect\citename{Sarna et al.\,}1995]{sdmm95} Sarna\,M.\,J.,
  Dhillon\,V.\,S., Marsh\,T.\,R., Marks\,P.\,B., 1995, MNRAS, L41

\bibitem[\protect\citename{Scargle\,}1982]{s82} Scargle\,J.\,D., 1982,
  ApJ, 263, 835

\bibitem[\protect\citename{Wood \&\ Saffer\,}1999]{w99} Wood\,J.\,H.,
  Saffer\,R., 1999, MNRAS, 305, 820
  
\bibitem[\protect\citename{Webbink\,}1984]{w84} Webbink\,R\,.F., 1984,
  ApJ, 277, 355

\end{thebibliography}
\end{document}